\documentclass[final,5p,times,twocolumn]{elsarticle}
\usepackage{amsmath,amsfonts}
\usepackage{algorithmic}
\usepackage[section]{placeins}
\usepackage{graphicx}
 
\usepackage{algorithm}
 
\usepackage{flushend}
\usepackage{amssymb}
\usepackage{color}
\usepackage{makecell}
\usepackage{flushend}
\usepackage{booktabs}
\usepackage{balance}
\usepackage[table,xcdraw]{xcolor}
\usepackage[normalem]{ulem}
\useunder{\uline}{\ul}{}
\usepackage{multirow}
\usepackage{amssymb}
\usepackage[table,xcdraw]{xcolor}

\journal{*}

\begin{document}

\begin{frontmatter}
\title{Self-supervised Fetal MRI 3D Reconstruction \\Based on Radiation Diffusion Generation Model}

\author[lable1]{Junpeng Tan}
\author[lable1,lable3]{Xin Zhang}
\author[lable1]{Yao Lv}
\author[lable2,lable3]{Xiangmin Xu}
\author[lable4]{Gang Li}

\affiliation[lable1]{organization={School of Electronics and Information, South China University of Technology},
            city={Guangzhou},
            postcode={510641}, 
            state={Guangdong},
            country={China}}
\affiliation[lable2]{organization={School of Future Technology, South China University of Technology},
            city={Guangzhou},
            postcode={511442}, 
            state={Guangdong},
            country={China}}
\affiliation[lable3]{organization={Pazhou Lab},
            city={Guangzhou},
            postcode={510330}, 
            state={Guangdong},
            country={China}} 
\affiliation[lable4]{organization={Department of Radiology, University of North Carolina at Chapel Hill},
            city={Chapel Hill},
            postcode={27599}, 
            state={North Carolina},
            country={The United States of America}}
\begin{abstract}
During fetal MRI acquisition, the unavoidable fetal motion and the non-static environment of the maternal uterus cause severe motion artifacts and noise. Although the use of multiple stacks can handle slice-to-volume motion correction and artifact removal problems, there are still several problems: 1) The slice-to-volume method usually uses slices as input, which cannot well solve the problem of uniform intensity distribution and complementarity in regions of different fetal MRI stacks; 2) The integrity of 3D space is not considered, which adversely affects the discrimination and generation of globally consistent information in fetal MRI; 3) Fetal MRI with severe motion artifacts in the real-world cannot achieve high-quality super-resolution reconstruction. To address these issues, we propose a novel fetal brain MRI high-quality volume reconstruction method, called the Radiation Diffusion Generation Model (RDGM). It is a self-supervised generation method, which incorporates the idea of Neural Radiation Field (NeRF) based on the coordinate generation and diffusion model based on super-resolution generation. To solve regional intensity heterogeneity in different directions, we use a pre-trained transformer model for slice registration, and then, a new regionally Consistent Implicit Neural Representation (CINR) network sub-module is proposed. CINR can generate the initial volume by combining a coordinate association map of two different coordinate mapping spaces. To enhance volume global consistency and discrimination, we introduce the Volume Diffusion Super-resolution Generation (VDSG) mechanism. The global intensity discriminant generation from volume-to-volume is carried out using the idea of diffusion generation, and CINR becomes the deviation intensity generation network of the volume-to-volume diffusion model. Finally, the experimental results on real-world fetal brain MRI stacks demonstrate the state-of-the-art performance of our method.
\end{abstract}




\end{frontmatter}


\section{Introduction}
\label{}

Fetal Magnetic Resonance Imaging (MRI) is an important means to study and monitor fetal brain development~\cite{khawam2021fetal}\cite{hughes2017dedicated}. The non-static states of the fetus and maternal uterus typically cause serious motion artifacts and other uncertain noises when acquiring the fetal MRI. Even with fast 2D sequences, fetal MRI remains susceptible to inter-slice motion artifacts~\cite{kuklisova2012reconstruction}, resulting in the misalignment of slices in the same stack. To this end, some dedicated image processing methods have been proposed, such as motion correction~\cite{godenschweger2016motion}, and super-resolution reconstruction~\cite{de2022simulation}. Although these methods can reduce the motion artifacts of slices, there are still some problems with 3D space continuity such as insufficient inter-slice motion correction and noisy images. 

Therefore, the fetal brain 3D volume reconstruction using multiple fetal MRI stacks has received increasing attention. In the early work, Ali et al. proposed a robust M-estimation solution that minimizes the robust error norm function between slices generated by the image requisition model and real slices~\cite{gholipour2010robust}. Maria et al. proposed a method to reconstruct fetal volume MRI from 2D slices~\cite{marami2016motion}, by combining intensity matching and robust statistics from 2D slices and excluding the misregistered or damaged slices. Those Slice-to-Volume (SVR) methods are widely used for fetal brain motion correction after rigid registration only. However, rigid registration methods cannot better solve the misregistration problem caused by the fetal brain deformation movement. To solve it, non-rigid motion correction deformable SVR methods are proposed. Like, Amir et al. proposed Patch-to-Volume Reconstruction (PVR)~\cite{alansary2017pvr}, which could carry out non-rigid reconstruction of fetal MRI structures with deformation in slices. Alena et al. proposed non-rigid motion correction deformable SVR (DSVR)~\cite{uus2020deformable}. Compared with rigid registration SVR methods and non-rigid motion correction deformable SVR methods, the reconstruction performance of non-rigid motion correction deformable SVR methods is improved. Notably, these SVR methods are all refactoring methods based on traditional methods, and the processing steps are very tedious.

\begin{figure*}
\centering
\includegraphics[width=0.9\textwidth]{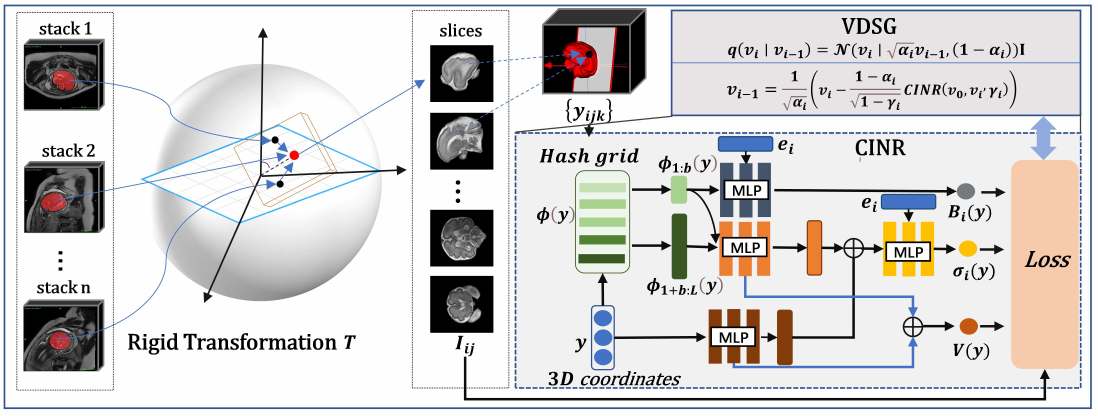}
\caption{The general block diagram of the method RDGM is presented. We take multiple stacks of fetal MRI and corresponding masks as input and use pre-trained model SVoRT for rigid registration of slices. Then SVR is performed with the proposed CINR. Finally, the VDSG mechanism is introduced to generate the super-resolution fetal brain MRI 3D volume.} \label{fig1}
\end{figure*}

To further improve the fetal brain MRI volume reconstruction quality and reduce processing steps, some deep learning-based methods have been proposed. Xu et al. proposed Spatio-Temporal Resolution Enhancement with Simulated Scans (STRESS)~\cite{xu2021stress}, a self-supervised super-resolution framework. Deep learning-based methods use simulated motion-corrupted slices of fetal MRI to train models, with the expectation of achieving 3D volume reconstruction of real fetal MRI. Xu et al. proposed Iterative Transformer for Slice-to-Volume Registration (SVoRT)~\cite{xu2022svort}, this method could realize slice-to-volume registration and 3D volume reconstruction. Shi et al. proposed an Affinity Fusion-based Framework for Iteratively Random Motion correction (AFFIRM)~\cite{shi2022affirm}, which iteratively estimated the slice motion and used the scattered data approximation technology to achieve fetal 3D volume reconstruction. However, Due to the lack of real labels in the acquired low-quality fetal MRI data, it is difficult to meet the needs in achieving real-world low-quality fetal 3D volume reconstruction. To this end, Xu et al. proposed implicit Neural Representation for SVR (NeSVoR)~\cite{xu2023nesvor}, which was a self-supervised implicit network~\cite{mildenhall2021nerf} and overcame the high storage costs of dense discretized voxel grids~\cite{tancik2020fourier}. Although NeSVoR could process real-world low-quality fetal 3D volume reconstruction, there are still defects in fetal MRI with severe motion artifacts and noise. Inter-slice regional consistency, as well as the overall integrity of the volume, are not considered.

To address these issues, we propose a novel fetal brain MRI 3D volume reconstruction method, called Radiation Diffusion Generation Model (RDGM). The main contributions are as follows: 1) We propose a new self-supervised framework for fetal 3D volume reconstruction. RDGM introduces the ideas of Nerve Radiation Field (NeRF) and diffusion model to achieve slices rigid registration, SVR, and volume-to-volume super-resolution reconstruction. 2) The new Consistency Implicit Neural Representation (CINR) network is proposed to generate an initial fetal brain MRI 3D volume. It can enhance the consistency association of regional coordinates intensity by combining the coordinate association map of two different coordinate mapping spaces. 3) To further improve the initial fetal brain MRI 3D volume quality, the volume Diffusion Super-resolution Generation mechanism (VDSG) mechanism is proposed. In the VDSG step, CINR is also used as the deviation-generated score network in the diffusion model to realize global super-resolution generation. 

Finally, CINR enhances the regional consistency of SVR, and VDSG improves the overall strength discrimination of volume reconstruction. RDGM can achieve high-quality reconstruction in real-world fetal MRI with severe artifacts.

The rest sections of this study are organized as follows. Section 2 gives a brief overview of related work. The proposed method is described in detail in Section 3. Section 4 reports extensive experimental results and discussion, followed by conclusions and future work in Section 5.

\section{Related Works}
Fetal MRI is usually affected by noise, slice motion estimation errors, and slice intensity inhomogeneity. There is an urgent need to address these effects on fetal MRI volume reconstruction quality. Therefore, some research has been done to deal with fetal MRI high-quality reconstruction problems. In this section, we expand our review on the following three aspects: 1) The general observation models of fetal brain 2D slices; 2) 3D vision reconstruction of Neural Radiance Field methods (NeRF); and 3) super-resolution generative diffusion models.

\subsection{Generic Observation Models}

According to existing studies, the most image acquisition model in the context of fetal MRI is the linear model. During data acquisition, the MRI Scanner acquires multiple stacks of low-resolution 2D slices, which are downsampled, degraded, and aliased with the high-resolution original scene. Like, \cite{rousseau2006registration} \cite{rousseau2010super}, and Fogtmann et al. \cite{fogtmann2013unified} demonstrated a typical correction model including motion correction, deblurring, and upsampling processes. Usually, during fetal MRI data acquisition, slice thickness needs to be taken into account. The 3D Point-Spread-Function (PSF) ~\cite{zhang2007gaussian} is a good approximation of a 3D Gaussian function with full-width half of the maximum value equal to the sum of slice thickness in the chosen slice direction 1.2$\times$ voxel size plane \cite{kuklisova2012reconstruction}. However, due to the presence of noise and the insufficient number of acquired low-resolution images, the above inverse problem is considered ill-posed, which means that it usually has no meaningful solution \cite{rousseau2013btk}. A natural solution to this problem is to add prior information constraints about fetal MRI, that is, to exploit known properties that fetal MRI possesses, such as motion estimation and intensity smoothness.

The motion estimation problem is solved by image registration, which mainly compensates for the motion occurring between slices of low-resolution images, usually generated by the short and fast motion of the fetal head \cite{andersson2017towards}. Some existing methods \cite{ferrante2017slice} \cite{makropoulos2018review}  \cite{uus2022retrospective} addressed this issue using a voxel-based registration approach, where the fetal motion was modeled as a full 6-DOF transformation (three translations and three rotations). Typically, they consist of (1) The non-rigid deformable of the low-resolution slice (PVR); (2) The global co-registering of the low-resolution slice (volume-to-volume registration); (3) The hierarchical alignment of each low-resolution slice (SVR). 

Besides, Interpolating super-resolution model for achieving intensity smoothness in high-quality fetal MRI \cite{mcdonagh2017context}. They include performing iterative reconstruction procedures that interleave rigid joint registration of low-resolution images and Scatter Data Interpolation (SDI). After estimating the motion parameters, a bias field correction step is performed to correct the local relative intensity distortion between the low-resolution MRI. Finally, a computationally efficient local neighborhood Gaussian kernel SDI is used for reconstruction. However, SDI reduces the spatial frequency content and thus leads to excessive blurring. To this end, Sander et al. \cite{sander2022autoencoding} proposed semantically smooth interpolation in through-plane direction. The method exploited the latent space generated by autoencoders. Algorithms that introduce spatial regularization in the fetal MRI reconstruction task have emerged in subsequent studies \cite{tourbier2015efficient}, which fall into two main categories: (1) deterministic variational models and (2) stochastic Bayesian models. However, these general observation models are not good and fast in the reconstruction of low-quality fetal MRI data to reconstruct satisfactory fetal brain MRI. However, these general methods generally involve complex processing steps and consume some time. There are deficiencies in reconstructing low-quality fetal MRI data.

\subsection{Neural Radiance Fields}

Recently, implicit functions have been widely used by researchers to represent properties such as 3D geometry and the appearance of objects and scenes. These methods use the Multi-Layer Perceptron (MLP) as the function to map continuous spatial coordinates and viewpoints into the signed distance, occupancy, color, or density values using a hash spatial encoding function $\gamma(.)$. Advances in implicit representations and more recently in differentiable rendering \cite{jin2023tensoir} have led to NeRF \cite{mildenhall2021nerf}, which integrates view-varying image properties by storing volume density from multi-view observations and view-dependent transparency to render properties such as the color of an entire object or scene. Generally, we use the MLP $F(.)$ that maps scene coordinates $y$ and view orientation $d$ to their corresponding densities ${\sigma}$ and RGB color values ${c}$, conditioned on shape and appearance codes. The NeRF methods are generally expressed as:
\begin{equation}
({c}, {\sigma})=F\left(\gamma(y), \gamma({d}), {z}_{{a}}, {z}_{{s}}\right)
\end{equation}
where, ${z}_{{a}}$ and ${z}_{{s}}$
are the shape information and appearance information, respectively. 

Existing NeRF methods generally need to satisfy several assumptions \cite{mi2022im2nerf}: (1) the camera pose of the input image and its intrinsic image properties are known $\xi$, we sample pixel rays ${r}(t)={o}+t {d}$ by given their origin ${o}$ and direction ${d}$.
(2) the pixel densities and colors $({\sigma}_{i}; {c}_{i})$ of the sampled points along the ray are implicit representations optimized on a single static scene by MLP. (3) Many input images from multiple views are required to achieve optimal volumetric rendering results:
\begin{equation}
\begin{array}{r}
\hat{{c}}({r})=\sum_{i=1}^{N} T_{i}\left(1-\exp \left(-\sigma_{i} \delta_{i}\right)\right) c_{i} \\
\text { where } T_{i}=\exp \left(-\sum_{j=1}^{i-1} \sigma_{j} \delta_{j}\right)
\end{array}
\end{equation}
where $N$ denotes the number of evenly-spaced pixels between the bounds. $\delta_i=t_{i+1}-t_{i}$ is the distance of adjacent samples, $t_i \in [t_n,t_f]$ and $t_n$ and $t_f$ are the near and far bounds. More recently, the emergence of camera-estimated implicit neural representations allows NeRF to work with a wider range of data from complex scenes \cite{meng2021gnerf}. Especially, it has gained attention in the field of SLAM \cite{chung2023orbeez}. At the same time, a large number of view data collection is also a drawback restricting the development of the NeRF method \cite{xu2022sinnerf} \cite{deng2023nerdi}. Therefore, how to achieve faster and high-quality 3D reconstruction of complex appearances or scenes from a small amount of input view data has attracted much attention.

\subsection{Diffusion Models}

To our knowledge, diffusion models \cite{rombach2022high} are emerging image generation models that can generate high-quality and high-fidelity image content. The diffusion model in its basic form \cite{chen2023seeing} is a probabilistic model defined by a bidirectional state Markov chain. First, the forward diffusion process gradually adds Gaussian noise to the origin data until it destroys the isotropy of the data. (2) The back-propagation process learns a model with network parameter $\theta$ distribution $p_\theta(x)$ by modeling the posterior distribution $q(x)$ in each state and finally obtains a denoising network training process from the original data score \cite{croitoru2023diffusion}. Formally, Diffusion models are generative models with a Markov Chain structure, suppose it has a fixed length for $t$ Markov Chain $x_t \to x_{t-1} \to \dots \to x_1 \to x_0$, which has the following joint distribution:
\begin{equation}
\begin{array}{r}
p_{\theta}\left({x}_{0: t}\right)=p_{\theta}^{(t)}\left({x}_{t}\right) \prod_{i=0}^{t-1} p_{\theta}^{(i)}\left({x}_{i} \mid {x}_{i+1}\right).
\end{array}
\end{equation}

After drawing $x_{0:t}$, only $x_0$ is kept as the sample of the generative model. The reverse can be expressed as conditional probability $q(x_{i - 1} | x_i)$, $i = 1,..., t$ and $x_i$ are obtained by corrupting the image $x_{i - 1}$ with Gaussian noise. To train a diffusion model, a fixed, factorized variational inference distribution is introduced:
\begin{equation}
\begin{array}{r}
q\left({x}_{1: t} \mid {x}_{0}\right)=q^{(t)}\left({x}_{t} \mid {x}_{0}\right) \prod_{i=0}^{t-1} q^{(i)}\left({x}_{i} \mid {x}_{i+1}, {x}_{0}\right).
\end{array}
\end{equation}
This leads to the Evidence Lower Bound (ELB) for the maximum likelihood objective \cite{kawar2022denoising}. The ELB can be simplified to the noise autoencoder objective function:
\begin{equation}
\begin{array}{r}
\sum_{i=1}^{t} \gamma_{i} \mathbb{E}_{\left({x}_{0}, {x}_{i}\right) \sim q\left({x}_{0}\right) q\left({x}_{i} \mid {x}_{0}\right)}\left[\left\|{x}_{0}-f_{\theta}^{(i)}\left({x}_{i}\right)\right\|_{2}^{2}\right],
\end{array}
\end{equation}
where $f_\theta(i)$ is a $\theta$-parameterized neural network that aims to recover a noiseless observation from a noisy $x_i$, and $ \gamma_{1:t}$ are a set of coefficients that depend on $q\left({x}_{1:t} \mid {x}_{0}\right)$. Usually, the acquisition of a blurry fetal MRI is an ill-posed inverse problem. Diffusion models aim to estimate one (or several) high-quality images from a blurry observation. The same applies to low-quality fetal MRI super-resolution reconstruction.

This paper aims to achieve rapid and high-quality fetal brain MRI reconstruction in low-quality fetal MRI data. We will abandon the generic observation model methods. Then, the multi-view reconstruction of NeRF and the noise reduction reconstruction of diffusion super-resolution are combined to achieve high-quality and high-fidelity fetal brain MRI reconstruction.

\section{Methods}
As shown in Fig.~\ref{fig1}, multiple stacks of fetal MRI are donated as $X=\left[x_{1}, x_{2}, \cdots, x_{n}\right]$, and $n$ is the number of stacks. First, NiftyMIC~\cite{ebner2020automated} is used to segment $X$, and obtains the masks for each stack $M=\left[m_{1}, m_{1}, \cdots, m_{n}\right]$. Then, we use a pre-trained model to achieve rigid registration of each slice in $X$. The corresponding slice transformations can be expressed as $T=\left[T_{1}, T_{2}, \cdots, T_{r}\right]$ and the slices of all stacks denote as $Y=\left[y_{1}, y_{2}, \cdots, y_{r}\right]$, where $r$ is the number of all slices. In the CINR sub-module, let $I\in \mathbb {R}^{r \times N_{p}}$ represent the data of the acquired slices, $I_{ij}$ denotes the intensity of the $j$-th pixel in the $i$-th slice, and $N_{p}$ is the number of pixels in each slice. $V$ is an unknown reconstruction volume. All input slices can be represented as an array of all voxel coordinates as $Y\in \mathbb {R}^{N_{v} \times 3}$ in the CINR intensity generation network, and $N_{v}$ is the number of reconstruction volume voxels.

\subsection{Consistency Implicit Neural Representation Network}
Recently, the Implicit Neural Representation network (INR) has been widely used in 3D rendering, which uses the coordinate of pixels to generate pixel values~\cite{wu2021irem}\cite{martin2021nerf}. Especially. NeSVoR~\cite{xu2023nesvor}  used a hash grid encoding method~\cite{muller2022instant} to discretize the slice coordinates with multi-vertex trilinear interpolation so that the slice coordinates match a high-dimensional feature vector. Such spatial mapping would lose the regional correlation consistency of the original spatial coordinates. To this end, in this sub-module, we propose a new INR network, called the Consistency Implicit Neural Representation (CINR). Specifically, we introduce a regional intensity learning module for slice coordinates in two different coordinate mapping spaces, which is as follows:
\begin{equation}
\left[V_{1}(y), Z_{1}(y)\right]=M L P_{V_{1}}(y)
\end{equation}
\begin{equation}
\left[V_{2}(y), Z_{2}(y)\right]=M L P_{V_{2}}(\phi(y))
\end{equation}
\begin{equation}
V(y)=V_{1}(y)+V_{2}(y) ; Z(y)=Z_{1}(y)+Z_{2}(y)
\end{equation}
\noindent where $M L P_{V_{1}}(\cdot)$ and $M L P_{V_{2}}(\cdot)$ are the original regional intensity learning network and hash grid encoding intensity learning network, respectively. All $MLPs$ have one hidden layer with 64 units and $ReLU$ activation. $\phi(\cdot)$ is the hash grid encoding, $\phi(y)=\left[\phi_{1}(y), \phi_{2}(y), \cdots, \phi_{L}(y)\right]$ and $L$ is the length of hash gird encoding. $V(y)$ and $Z(y)$ are the coordinate feature vector and intensity feature vector at position $y$, respectively.

Further, we adopt a continuous slices acquisition model~\cite{xu2023nesvor} to generate the intensity of $j$-th pixel in the $i$-th slice $I_{ij}$.
\begin{equation}
I_{i j}=C_{i} \int_{\Omega} M_{i j}(y) B_{i}(y)\left[V(y)+\epsilon_{i}(y)\right] \mathrm{d} y \label{Eq.(4)}
\end{equation}
\noindent where $\Omega$ is the 3D mask region, $C_{i}$ is scaling factor, $M_{ij}(y)$ is the coefficient of spatially aligned as:

\begin{equation}
M_{ij}(y)=g(T_i^{-1}\circ y-p_{ij};\Sigma)
\end{equation}
\begin{equation}
g(u;\Sigma)=\frac{1}{\sqrt{(2\pi)^3\text{det}(\Sigma)}}\text{exp}(-\frac{1}{2}u^T\Sigma^{-1}u)
\end{equation}
where $T_{i}$ is the rigid transformation of the $i$-th slice from transformer-based pre-trained SVoRT. $p_{ij}$ is the location of pixel $I_{ij}$ in the slice coordinates. $\Sigma$ is the covariance matrix of the Gaussian PSF, $\Sigma=diag((\frac{1.2r_1}{2.355})^2,(\frac{1.2r_2}{2.355})^2,(\frac{r_3}{2.355})^2)$, where $r_{1}$ and $r_{2}$ are the in-plane pixel space and $r_{3}$ is the slice thickness. As shown in Eq. (12), $B_{i}(y)$ is the bias field, it is a smoothly varying function of spatial location. We use the low-level encoding $\phi_{1:b}(y)=[\dot{\phi_1(y)},\dot{\phi_2(y)},\cdots,\phi_b(y)]$ and the slice embedding $e_{i}$ to train $B_{i}(y)$.
\begin{equation}
B_i(y)=MLP_B(\phi_{1:b}(y),e_i)
\end{equation}
where $M L P_{B}(\cdot)$ is the bias filed training network. In Eq. (9), $\epsilon_{i}(y)$ is white Gaussian noise with $\quad\text{E}[\epsilon_i(y)]=0$ and $\quad\text{E}[\epsilon_i(y)\epsilon_i(z)]=\sigma_i^2(y)\delta(y-z)$, $\delta(\cdot)$ is the Dirac delta function. According to Eq. (9), the mean and variance of pixel $I_{ij}$ can be denoted as:
\begin{equation}
\overline{I}_{ij}=\mathbb{E}[I_{ij}]=C_i\int_\Omega M_{ij}(y)B_i(y)V(y)\mathrm{d}y
\end{equation}
\begin{equation}
\sigma_{ij}^2=\operatorname{var}\left(I_{ij}\right)=C_i^2\int_\Omega M_{ij}^2(y)B_i^2(y)\sigma_i^2(y)\mathrm{d}y
\end{equation}
where $\sigma_{i}^{2}(y)$ is variance by network training, which is slice-dependent. We adopt $M L P_{\sigma}(\cdot)$ to learn variance at location $y$ by combining the feature vector $Z(y)$ with the slice embedding $e_{i}$.
\begin{equation}
\sigma_i^2(y)=MLP_\sigma\left(Z(y),e_i\right)
\end{equation}
According to the continuous iterations of the training network $M L P_{V_{1}}(\cdot)$, $M L P_{V_{2}}(\cdot)$, $M L P_{B}(\cdot)$ and $M L P_{\sigma}(\cdot)$, variables $V(y)$, $M_{ij} (y)$, $B_{i} (y)$ and $\sigma_{ij}^{2}$ can be obtained. Combined with Eq. (4), the intensity of each pixel is calculated.

\subsection{Volume Diffusion Super-resolution Generation Mechanism}
Diffusion models~\cite{chung2022parallel}\cite{song2023pseudoinverse} have achieved good results in the fields of computer vision~\cite{kim2022diffusionclip}\cite{liu2022compositional}. Here, to improve the quality of fetal brain MRI, it is necessary to enhance volume global consistency and discrimination. We adopt the idea of diffusion generation for the volume-to-volume reconstruction of fetal brain MRI 3D volumes. Combined with the CINR of SVR, we propose a VDSG mechanism. It can learn the global feature intensity, and strengthen the uniform distribution of the overall intensity. 

In this sub-model, let the CINR output $V$ as the initial input of VDSG ($v_{0}$). We first define a forward Markovian diffusion process $q$ that gradually adds Gaussian noise to initial MRI $v_{0}$ over $t$ iterations:
\begin{equation}
q\left(\boldsymbol{v}_{1: t} \mid \boldsymbol{v}_{0}\right)=\prod_{i=1}^{t} q\left(\boldsymbol{v}_{i} \mid \boldsymbol{v}_{i-1}\right)
\end{equation}
\begin{equation}
q\left(\boldsymbol{v}_{i} \mid \boldsymbol{v}_{i-1}\right)=\mathcal{N}\left(\boldsymbol{v}_{i} \mid \sqrt{\alpha_{i}} \boldsymbol{v}_{i-1},\left(1-\alpha_{i}\right) \mathbf{I}\right)
\end{equation}
where the scalar parameters $\alpha_{1:t}$ are hyper-parameters, $\textbf{I}$ is unit tensor and $0<\alpha_{i}<1$. The distribution of $v_{i}$ given $v_{0}$ by marginalizing out the intermediate steps can be as

\begin{equation}
\begin{aligned}q(\boldsymbol{v}_i\mid\boldsymbol{v}_0)=\mathcal{N}(\boldsymbol{v}_i\mid\sqrt{\gamma_i}\boldsymbol{v}_0,(1-\gamma_i)\text{I})\end{aligned}
\end{equation}
where $\gamma_i=\prod_{m=1}^i\alpha_m $, recall that the generation model is trained to estimate noise. Here, instead of the Unet score network, we use CINR as the score generation network model. Furthermore, with some algebraic manipulation and completing the square, we can derive the posterior distribution of $v_{i-1}$ as
\begin{equation}
\boldsymbol{v}_{i-1}=\frac{1}{\sqrt{\alpha_i}}(\boldsymbol{v}_i-\frac{1-\alpha_i}{\sqrt{1-\gamma_i}}CINR(\boldsymbol{v}_0,\boldsymbol{v}_i,\gamma_i))
\end{equation}
Thus, given $v_{i}$, through the continuous iterative generation of Eq. (19), we approximate $v_{0}$ as
\begin{equation}
\widehat{\boldsymbol{v}}_0=\frac{1}{\sqrt{\gamma_i}}(\boldsymbol{v}_i-\sqrt{1-\gamma_i}CINR(\boldsymbol{v}_0,\boldsymbol{v}_i,\gamma_i))\label{Eq.(15)}
\end{equation}
Combined with CINR output $V$, we can obtain a high-quality fetal brain MRI volume by $V+\widehat{\boldsymbol{v}}_0$.
\subsection{Loss Functions}
Due to the CINR sub-model being a self-supervised training network, we introduce slice reconstruction loss, image regulation loss, and bias field loss.

\paragraph{\bfseries 1) Slice Reconstruction Loss:} We use the generated pixel intensity mean $\overline{I}_{i j}$ and variance $\sigma_{ij}^{2}$ to reconstruct the underlying volume by minimizing the negative log-likelihood Gaussian distribution. The function of this loss function is to optimize the regional consistency (a batch of data $\mathcal{B}\subset\{1,\cdots, N_s\}\times\{1,\cdots, N_p\}$) pixel intensity from the perspective of outlier removal. 
\begin{equation}
\mathcal{L}_I=\frac{1}{|\mathcal{B}|}\sum_{(i,j)\in \mathcal{B}}\lambda\frac{(I_{ij}-\bar{I}_{ij})^2}{2\sigma_{ij}^2}+\frac{1}{2}\log\left(\sigma_{ij}^2\right)
\end{equation}
where $I_{ij}$ is initial pixel intensity. $\lambda$ is the trade-off parameter.
\paragraph{\bfseries 2)Image Regulation Loss:} We adopt serval regularization methods $r_m(\cdot)$ to improve image quality and suppress noise. Here, we use the identity function (isotropic total variation), square function (first-order Tikhonov), and Huber function (edge-preserving). Here, we introduce the regulation loss as follows: 
\begin{equation}
\mathcal{R}_V=\sum_{m=1}^3\frac{2}{K|\mathcal {B}|}\sum_{(i,j)\in\mathcal{B}}\sum_{k=1}^{K/2}r_m(\frac{|V(x_{ijk})-V(x_{ijl})|}{\|x_{ijk}-x_{ijl}\|_2})
\end{equation}
Here, $K$ is the number of subjects, we split the $K$ into $K⁄2$ pairs. $\big|V\big(x_{ijk}\big)-V\big(x_{ijl}\big)\big|/\big\|x_{ijk}-x_{ijl}\big\|_2$ is the directional derivative for each pair, and $l=k+K/2$.
\paragraph{\bfseries 3) Bias Field Loss:} According to ~\cite{xu2023nesvor}, the bias field $B_{i} $ 
and volume $V$ is only unique up to a constant factor. To disambiguate $\left(B_i, V\right)\quad\text{}$and $(cB_i,\frac{1}{c}V)$, we should force the mean log bias field to be zero, where $c$ is any constant. Therefore, we can get the bias field loss function as follows:
\begin{equation}
\quad\mathcal R_B=\left(\frac{1}{K|\mathcal {B}|}\sum_{(i,j)\in\mathcal B}\sum_{k=1}^K\log B_i(x_{ijk})\right)^2
\end{equation}
Finally, the training optimization problem can be denoted as 
\begin{equation}
\underset{\theta}{arg min } \mathcal{L}(\theta),\quad\mathcal{L}=\mathcal{L}_I+\lambda_V\mathcal{R}_V+\lambda_B\mathcal{R}_B
\end{equation}
where parameters $\lambda_{V}$ and $\lambda_{B}$ are the trade-off weights for the regularization terms, $\theta$ is the set of trained network parameters.  The overall procedure of our proposed method is summarized in Algorithm 1.

\section{Experiments and Results}
\subsection{Datasets}
\begin{figure}[t]
\centering
\includegraphics[width=0.5\textwidth]{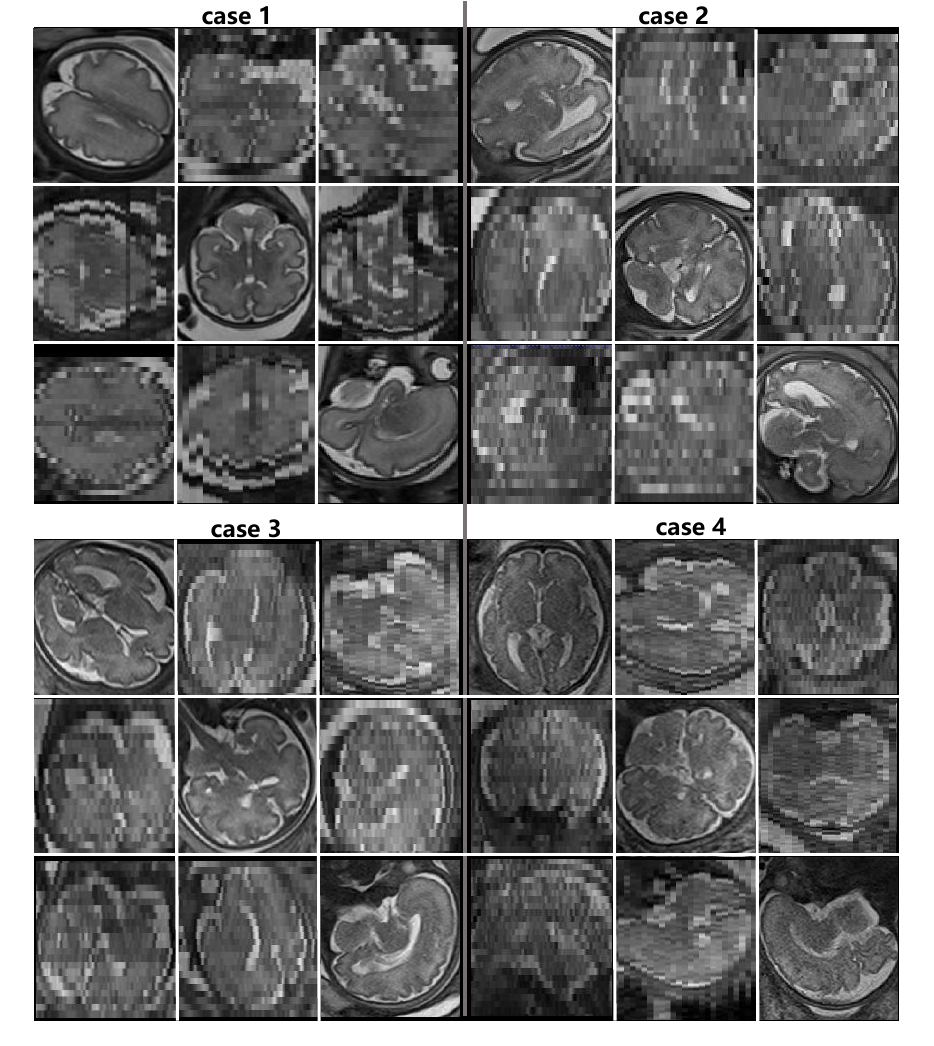}
\caption{Three real-world acquired fetal MRI subjects, where case 1 is from dataset A, case 2 is from dataset B, and case 3 and case 4 are from dataset C.} \label{fig2}
\end{figure}

We use three real-world datasets to validate the proposed model. The subjects of each dataset include at least three orthogonal views, such as axial, sagittal, and coronal orientations. Dataset A contains 132 stacks from 44 subjects, with the gestational age ranging from 24 week to 37 week. These stacks are acquired using a 3T  Philips Ingenia scanner and the resolution is $0.71\times0.71\times3.0$ $ mm^3$, $TR/TE=15000/177$ $ ms$. Dataset B contains 212 stacks from 46 subjects. These stacks are acquired using a 1.5T Siemens Avanto scanner and the resolution is $0.54\times0.54\times4.4$ $ mm^3$, $TR/TE=1350/92$ $ ms$. The gestational ages are from 21 week to 38 week. Dataset C contains 1106 stacks from 284 subjects. These stacks are acquired using a 1.5T Siemens Avanto scanner and the resolution is $0.54\times0.54\times4.4$ $ mm^3$, $TR/TE=1350/92$ $ ms$. The gestational ages are from 21 week to 36 week. Since our method is self-supervised training, it does not require a large amount of training data to learn the network parameters. As shown in Fig.~\ref{fig2}, we select several subjects from these datasets to verify the proposed model. Among them, case 1 is from dataset A with clear boundary structure and texture information. case 2 is from dataset B, which has a fuzzy fetal MRI with a small amount of noise on the data surface. case 3 and case 4 are from dataset C. Compared to datasets A and B, dataset C has worse acquisition quality, with large motion artifacts, blurred surface information, and large inter-slice spacing and motion. In particular, there is more noise in case 4. Some key information about datasets can be summarized in Table 1. 

\begin{algorithm}[H]
\caption{RDGM}\label{alg:alg0}
\begin{algorithmic}
\STATE \textbf{Input: }Multiple stacks of fetal MRI data $X=\left[x_{1}, x_{2}, \cdots, x_{n}\right]$, masks of each stack $M=\left[m_{1}, m_{1}, \cdots, m_{n}\right]$. The number of CINR training iterations is $T=5000$. 
\STATE 1: Initialize the parameters $batch_-size=4096$ and the number of samples for PSF during training $samples=256$. The learning rate is set as 0.001 and $\lambda=20$, $\lambda_V=2$, $\lambda_B=100$, $ \alpha_0=0.001$, $t=10$. 
\STATE 2: Extracting the region of interest utilizes $X=\left[x_{1}, x_{2}, \cdots, x_{n}\right]$ and $M=\left[m_{1}, m_{1}, \cdots, m_{n}\right]$. The rigid registration of slices $Y=\left[y_{1}, y_{2}, \cdots, y_{r}\right]$ uses a pre-trained model SVoRT.
\STATE 3: \textbf{while}
\STATE 4: \textbf{for} $iter<T$ \textbf{do} training CINR ($V$)
\STATE 5: Update the variables $\left[V_{1}(y), Z_{1}(y)\right]=M L P_{V_{1}}(y)$ by Eq. (6)
\STATE 6: Update the variables $\left[V_{2}(y), Z_{2}(y)\right]=M L P_{V_{2}}(\phi(y))$ by Eq. (7)
\STATE 7:Update the variables $V(y)=V_{1}(y)+V_{2}(y) ; Z(y)=Z_{1}(y)+Z_{2}(y)$ by Eq. (8)
\STATE 8: Compute the variables $B_i(y)=MLP_B(\phi_{1:b}(y),e_i)$ and $\sigma_i^2(y)=MLP_\sigma\left(Z(y),e_i\right)$ by Eq. (12) and Eq. (14)
\STATE 9: Obtain volume $V$ by generating the intensity of $j$-th pixel in the $i$-th slice $ I_{i j}=C_{i} \int_{\Omega} M_{i j}(y) B_{i}(y)\left[V(y)+\epsilon_{i}(y)\right] \mathrm{d} y$ by Eq. (9)
\STATE 10:  \textbf{end for}
\STATE 11: Gradually add Gaussian noise to initial MRI $V=v_{0}$ over $t$ iterations 
 $q\left(\boldsymbol{v}_{1: t} \mid \boldsymbol{v}_{0}\right)=\prod_{i=1}^{t} $ by Eq. (16)
\STATE 12: \textbf{for} $iter<t$ \textbf{do} computing VSCG ($\widehat{\boldsymbol{v}}_0$)
\STATE 13: Derive the posterior distribution of ${v}_{i-1}$\\
 $\boldsymbol{v}_{i-1}=\frac{1}{\sqrt{\alpha_i}}(\boldsymbol{v}_i-\frac{1-\alpha_i}{\sqrt{1-\gamma_i}}CINR(\boldsymbol{v}_0,\boldsymbol{v}_i,\gamma_i))$ by Eq. (19)
\STATE 14: \textbf{end for}
\STATE 15: \textbf{end while}
\STATE \textbf{output: $V+\widehat{\boldsymbol{v}}_0$}.
\end{algorithmic}
\label{alg0}
\end{algorithm}

\begin{table}[h]
\centering
\caption{Summary of Some Key Information about The Datasets.}\label{table1}
\resizebox{1\linewidth}{!}{
\begin{tabular}{c|cccc}
\hline
Datasets  & Subjects & Stacks & Resolution($mm^3$) & Gestational Age(week) \\
\hline
Dataset A & 44       & 132    & 0.71$\times$0.71$\times$3.0   & 24-37                 \\
Dataset B & 46       & 212    & 0.54$\times$0.54$\times$4.4   & 21-38                 \\
Dataset C & 284      & 1106   & 0.54$\times$0.54$\times$4.4   & 21-36                 \\
Atlas     & 1        & 3      & 0.8$\times$0.8$\times$0.8     & 26             \\
\hline
\end{tabular}}
\end{table}

In the visualization experiments below, we use 3 stacks for each subject to reconstruct volume so that every subject will contribute to the same number of stacks. The learning rate is 0.001 and $\lambda$ is set as 20, $\lambda_V=2$, $\lambda_B=100$, $ \alpha_0=0.001$, $t=10$. The experimental device is a GeForce RTX 3090 Ti 24G.

\subsection{Compared with Results of Different Methods}
\begin{figure*}[t]
\centering
\includegraphics[width=0.8\textwidth]{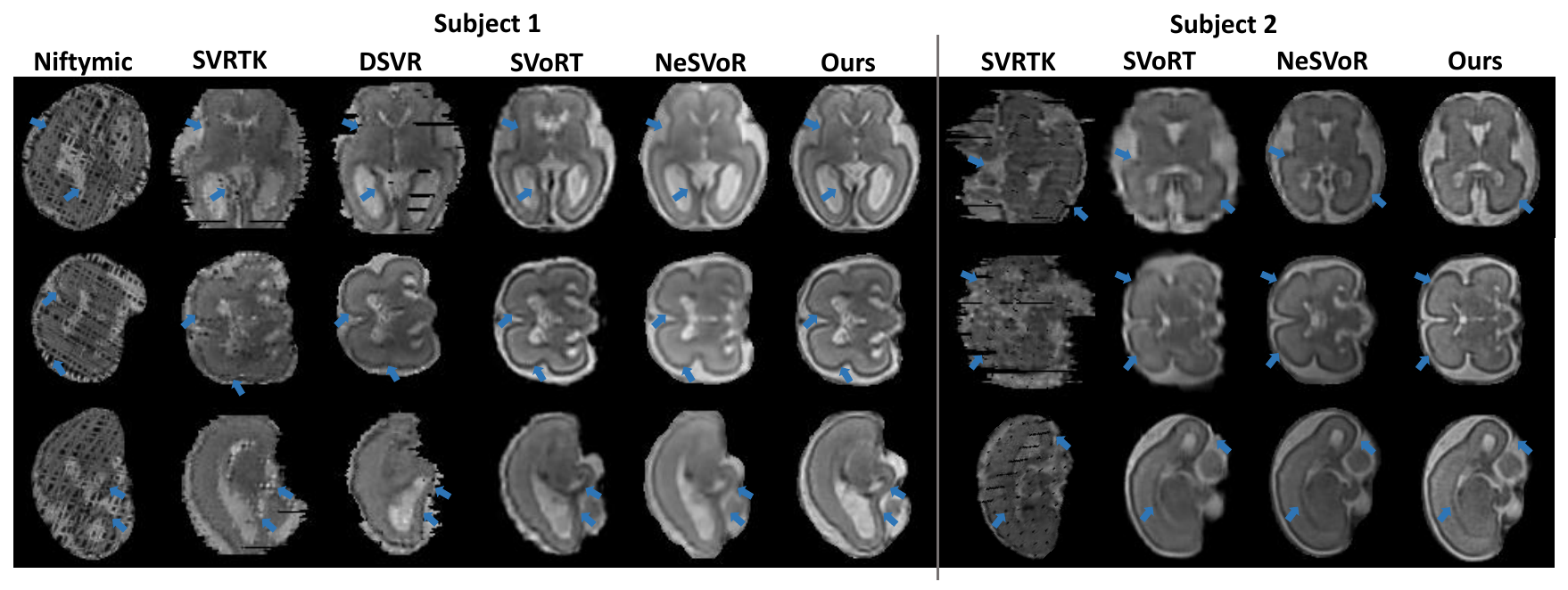}
\caption{The reconstruction volumes on different methods. Subjects 1 (25 week) and 2 (26 week) are from dataset A and dataset B, respectively.} \label{fig3}
\end{figure*}

We adopt some the state-of-the-art SVR methods as the baselines, which have open-source implementation: NiftyMic~\cite{ebner2020automated}, SVRTK~\cite{kuklisova2012reconstruction}, DSVR~\cite{uus2020deformable}, SVoRT~\cite{xu2022svort} and NeSVoR~\cite{xu2023nesvor}. In all methods, we use the fixed hyper-parameters that are proposed from papers or tuning the best results. As shown in Fig.~\ref{fig3}, our method achieves the best reconstruction volumes. Particularly, our method has more apparent textures in dataset A, without significant noise or artifacts compared with the baseline methods. Compared with the NeSVoR method, our proposed modules CINR and VDSG can improve the fetal brain MRI reconstruction volumes. However, the experimental results of subject 2 are slightly less accurate than those of subject 1 and have a little artifact. From Fig.~\ref{fig2}, dataset B has more severe motion artifacts. Due to severe motion in subject 2, the experimental results of NiftyMic and DSVR are not shown in Fig.~\ref{fig3}.

\begin{figure*}[t]
\centering
\includegraphics[width=0.8\textwidth]{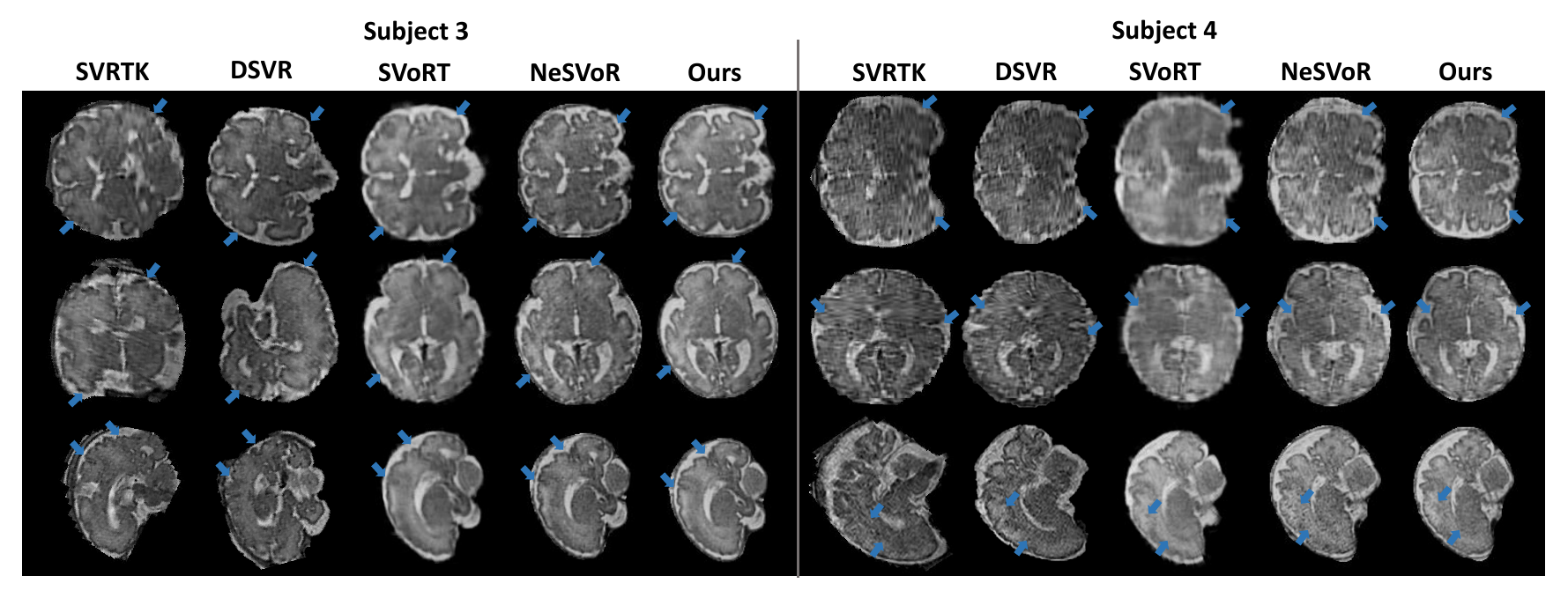}
\caption{The reconstruction volumes on different methods. Subject 3 (34 week) and subject 4 (35 week) are from dataset C.} \label{fig31}
\end{figure*}

In Fig.~\ref{fig31}, two samples from dataset C are presented through the results of different fetal MRI volume reconstruction methods. Compared with other state-of-the-art methods, our proposed method can achieve optimal reconstruction results. This indicates that our proposed method can achieve satisfactory reconstruction results even when the quality of the collected data is poor. This depends on the fact that we have used self-supervised training for region consistency and diffusion generation for super-resolution. Compared with NeSVoR, our method has more advantages in noise control, and can achieve uniform distribution of stool texture even in the case of poor acquisition quality. Observing the third-best result SVoRT, although it can reconstruct a relatively good structure of white matter and gray matter in subject 3, the reconstruction effect ratio in subject 4 is not very good. This indicates that this method cannot deal with low-quality fetal MRI data very well. Notably, we do not show the fetal MRI reconstruction results of NiftyMic compared to Fig.~\ref{fig31}. This is because the reconstruction quality of NiftyMic is worse than that of SVRTK and DSVR. It does not distinguish well between images from three directions. According to Fig.~\ref{fig3} comparison, NiftyMic also has the worst reconstruction results in high-quality dataset A.

\subsection{Simulated Fetal Brain Data}
\begin{figure}[t]
\centering
\includegraphics[width=0.3\textwidth]{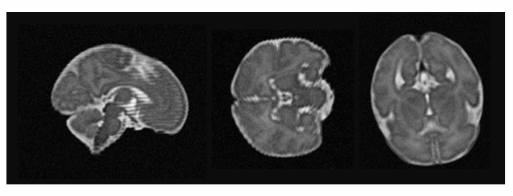}
\caption{A simulated stack of the fetal brain data (26 week) in the Fetal Brain Atlas dataset.} \label{figsim}
\end{figure}

Here, we further perform a quantitative analysis of the fetal volume reconstruction algorithm. Fig.~\ref{figsim} illustrates a simulated stack of the fetal brain data (26 weeks). In this experiment, the Fetal Brain Atlas dataset (26 week)~\cite{gholipour2017normative} is used to simulate fetal data~\cite{xu2020semi} reconstruction with 0.8 $mm$ isotropic resolution and match the resolution of the original ground truth. Specifically, the maximum translation and rotation motion in the motion trajectory dataset is 21.4 mm/s and 59.7 degree/s respectively. We compare the reconstructed volume and ground truth using various quantitative metrics, including Peak Signal-to-Noise Ratio (PSNR), Structural Similarity (SSIM), Root Mean Square Error (RMSE), and Normalized Cross-Correlation (NCC). As shown in Table 2, our proposed method achieves the best results in all quantitative metrics. RDGM improves by 0.32 dB on PSNR and 0.029 on SSIM compared with baseline. Especially, on quantitative metrics of RMSE and NCC, RDGM greatly improved compared with other methods.

\begin{table}[h]
\centering
\caption{Mean Values of Quantitative Metrics for Models on The Fetal Brain Simulated Data (Standard Deviation) and The Best Results Are Highlighted in Black.}\label{table1}
\resizebox{1\linewidth}{!}{
\begin{tabular}{c|c|c|c|c}
\hline
\bfseries \quad Methods\quad \ & \ \bfseries \quad PSNR(dB)$\uparrow$ \quad  & \bfseries SSIM$\uparrow$ &\bfseries RMSE$\downarrow$ &\bfseries NCC$\uparrow$\\
\hline
SVRTK & 23.7439(0.1234) &  0.3991(0.0012)	&0.8640(0.0012)	 &0.2785(0.0029)\\ \hline
 SVoRT & 24.8237(0.2865)	 &0.3906(0.0052)  &0.9581(0.0079)	 &0.1816(0.0555)\\  \hline
DSVR &25.5134(2.2824)  &0.3959(0.0012)  &0.8693(0.0013) &0.2728(0.0030)\\  \hline
NeSVoR &26.0515(0.0554)  & 0.3782(0.0005) & 0.8846(0.0009)	&0.4567(0.0041)\\ \hline
 Ours &\bfseries 26.3754(0.0126)& \bfseries0.4200(0.0001)	 &\bfseries0.7807(0.0001) &\bfseries0.5452(0.0029)\\ \hline
\end{tabular}}
\end{table}

\begin{figure}[t]
\centering
\includegraphics[width=0.5\textwidth]{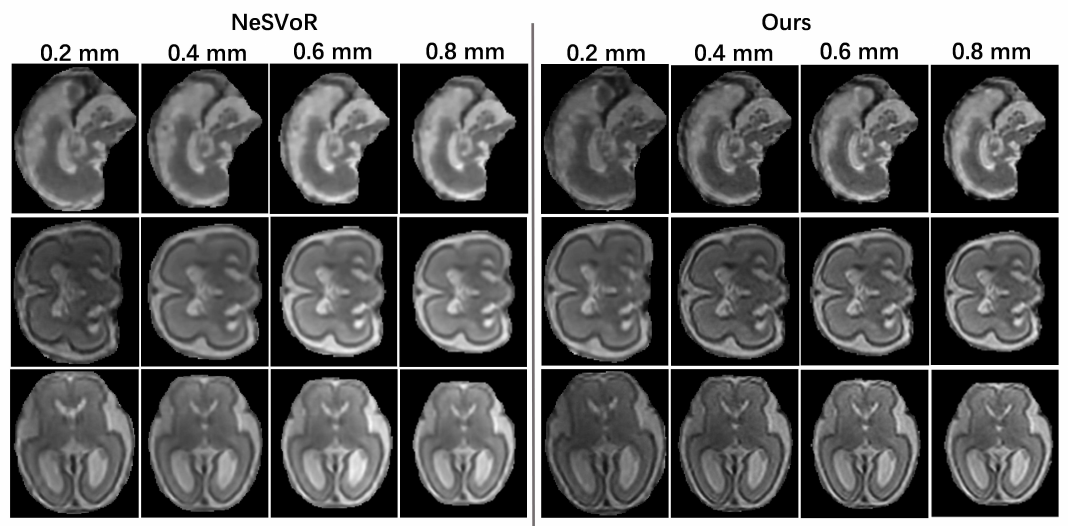}
\caption{The results of different resolution, the subject (25 week) is from dataset A.} \label{fig4}
\end{figure}

\subsection{Reconstructing Volumes at Different Resolutions}

To verify the effectiveness of our proposed method for fetal brain MRI reconstruction at different resolutions, Fig.~\ref{fig4} shows reconstruction volumes at different isotropic voxel spacing (0.8 $mm$, 0.6 $mm$, 0.4 $mm$, and 0.2 $mm$). We also perform the same experiment using the baseline method NeSVoR for comparison. Compared with NeSVoR, our proposed method is superior to it in all resolutions. NeSVoR reconstruction results are a little ambiguous, which makes it difficult to distinguish brain surface characteristics. Especially at 0.2 $mm$ reconstruction resolution, we can barely see the texture details of the brain. The specific reasons can be learned by the following ablation study. Meanwhile, compared with the texture of these two methods, the texture reconstructed by our method has less noise. This is because we introduced the VDSG mechanism to carry out super-resolution reconstruction of the reconstructed volume of CINR.

\subsection{Reconstructing Volumes at Different Gestational Ages}
\begin{figure*}[t]
\centering
\includegraphics[width=0.8\textwidth]{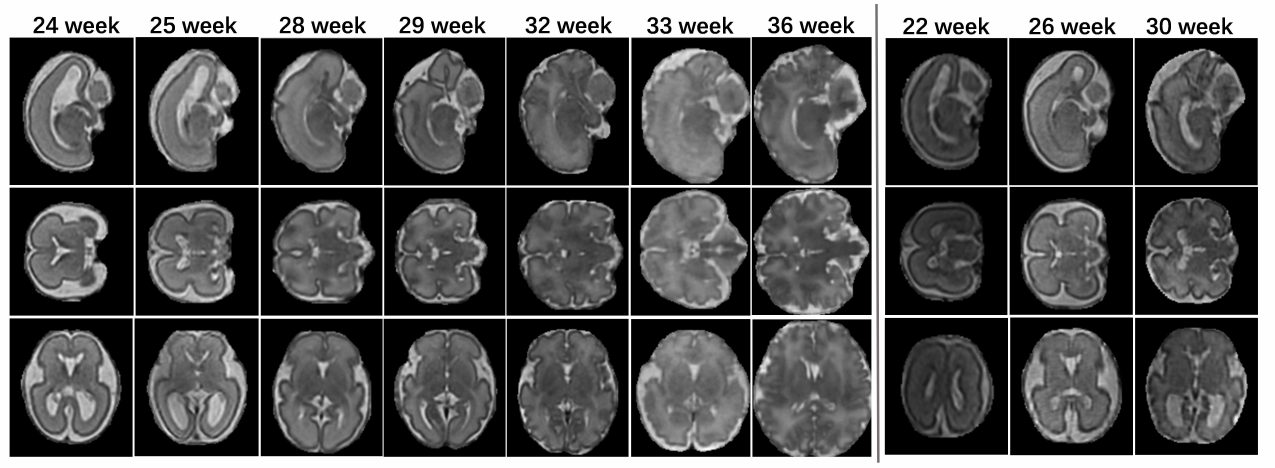}
\caption{The reconstruction results at different gestational ages. The left subjects are from dataset A, and the right subjects are from dataset B.} \label{fig5}
\end{figure*}

To verify the effectiveness of our method with more samples, we conduct MRI reconstruction experiments on fetuses of different gestational ages. The quality of fetal MRI collection in dataset A is better than that in datasets B and C. We will make extensive use of dataset A cases to demonstrate brain MRI at different gestational ages. At the same time, the collection quality of dataset B is better than that of dataset C. We also select a small number of cases of dataset B to show brain MRI after reconstruction at different gestational ages. As Fig.~\ref{fig5} shows, RDGM reconstructs great results at different gestational ages. We can see a tendency for fetal brains to grow with gestational age. Especially, in dataset A, the volume of subjects is clearly reconstructed, and the intensity of each position is evenly distributed. This is because the CINR sub-module encourages region-consistent pixel intensity. However, there is little noise in the reconstruction of the dataset B. According to Fig.~\ref{fig2}, the degree of motion artifact in dataset A is smaller than that in dataset B and there are unclear boundaries in each stack of input data. 

\begin{figure}[t]
\centering
\includegraphics[width=0.5\textwidth]{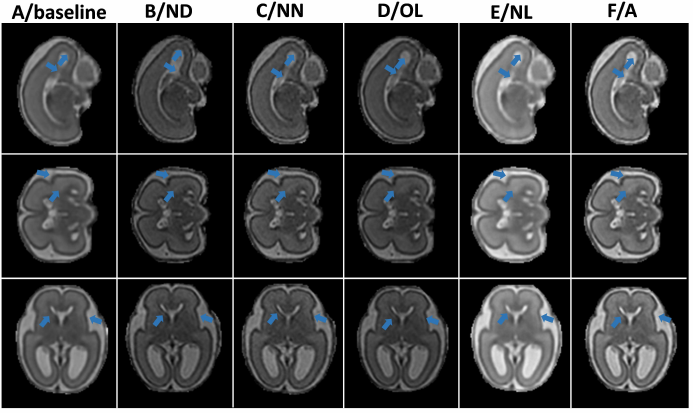}
\caption{The results of the ablation study, the subject is from dataset A (24 week). } \label{fig6}
\end{figure}

\subsection{Ablation Study}
To investigate the contributions of each innovative component in RDGM, we evaluate the model on the dataset A subject (24 week) by ablating the baseline (NeSVoR), the proposed method CINR without VDSG mechanism (ND), the proposed method VDSG without CINR (NN), only improving loss functions (OL), the proposed method without our improved loss function (NL), and RDGM. As shown in Fig.~\ref{fig6}, $A$ is the baseline (NeSVoR); $B$ is the ND; $C$ is the NN; $D$ is the OL. $E$ is the NL; $F$ is the proposed model RDGM. Compared with the baseline, the reconstruction results of all our improved modules are better than NeSVoR. Only by improving the regionalized consistent CINR network structure, the reconstructed volume pixel intensity is more uniform, and the boundary and texture are more distinguishable. When we introduce the VDSG mechanism, the surface texture noise is reduced. When only the improved loss function is introduced, the reconstructed volume can achieve clearer boundary information and reduce the amount of noise. Especially, when we add CINR and VDSG together without introducing new loss functions, the reconstruction results can achieve the heterogeneity of the global intensity of the brain structure and reduce noise pollution. However, the boundary information is lost, which shows that the improved loss functions are also critical.

\subsection{Analysis of Different Number Stack Reconstruction Results}

\begin{figure}[t]
\centering
\includegraphics[width=0.5\textwidth]{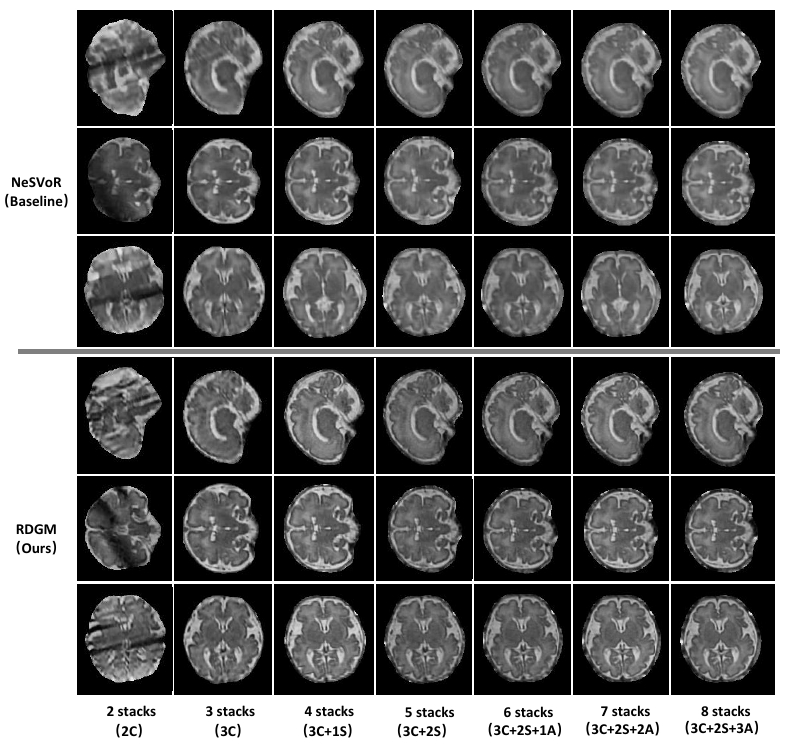}
\caption{Reconstruction results of different numbers of input stacks (2-8) randomly selected for subjects in dataset C. } \label{fig13}
\end{figure}

In this subsection, the reconstruction results using different number stacks will be analyzed. As shown in Fig.~\ref{fig13}, we show the reconstruction results for different numbers of input stacks, from 2 to 8, randomly selected by the subjects in dataset C. By comparing the baseline method NeSVoR with our proposed method RDGM, RDGM can achieve good reconstruction results for all different number of input stacks. Notably, when two stacks are input in the same direction, neither of these methods can achieve satisfactory reconstruction results. This indicates that the input data of this sample is of low quality. At the same time, both NeSVoR and RDGM can continuously improve the reconstruction quality as the number of input stacks increases. However, after the comparative analysis in Fig.~\ref{fig13}, our method has almost the same reconstruction effect when inputting 6 stacks as inputting 7 stacks and 8 stacks. This demonstrates that our method does not require a larger number of stacks than the baseline method NeSVoR for satisfying fetal brain MRI reconstruction when performing low-quality acquisition data reconstruction. This also proves that our proposed method is more competitive with the same number of stacks and a small number of stacks.

\begin{figure}[t]
\centering
\includegraphics[width=0.5\textwidth]{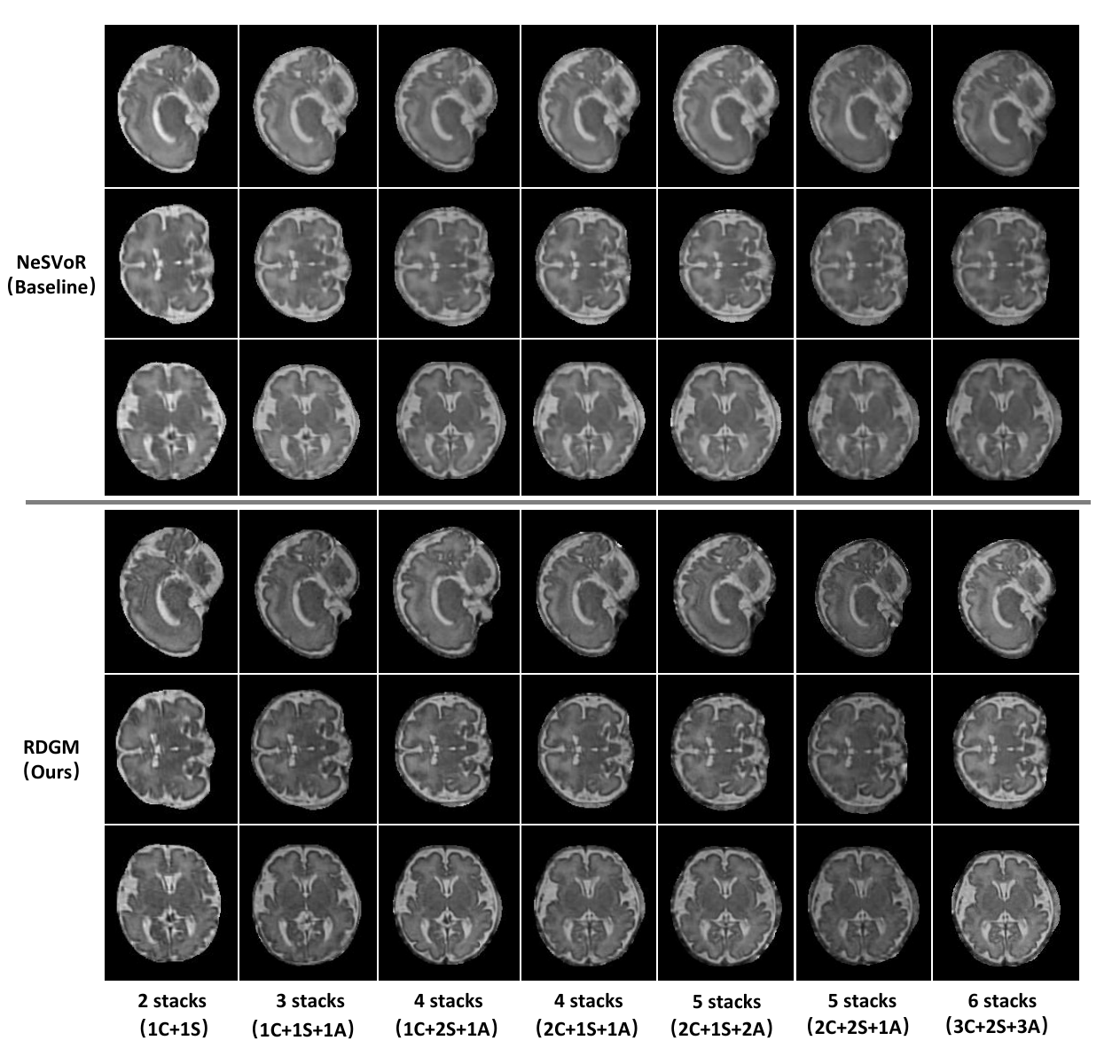}
\caption{Reconstruction results of different number of input stacks (2-6) selected for subjects in dataset C.} \label{fig12}
\end{figure}

In Fig.~\ref{fig13}, we find that there is still a deficiency in the reconstruction effect by randomly selecting stacks with the same orientation (2 and 3), especially when the number of inputs is 2 stacks. To this end, we need to further compare and analyze the differences between our proposed method and the baseline method. As Fig.~\ref{fig12} shows, we show the reconstruction results for different numbers of input stacks targeted. Each input stack must contain acquisition samples of different orientations. After the comparative analysis of Fig.~\ref{fig13} and Fig.~\ref{fig12}, the reconstruction results of Fig.~\ref{fig12} are better than those of Fig.~\ref{fig13}. This is because the input stacks from different directions play a key role in the reconstruction of fetal brain MRI, making the information and data structure of stacks from different directions complimentary. At the same time, the reconstructed results of our proposed method are also better than those of the baseline method.  Especially with a very small number of stack inputs, our proposed method can also reconstruct good fetal brain MRI 3D effects. This indicates that our proposed method has more advantages in the fusion of different methods stacks.

\subsection{Downstream Task Analysis}

\begin{figure}[t]
\centering
\includegraphics[width=0.5\textwidth]{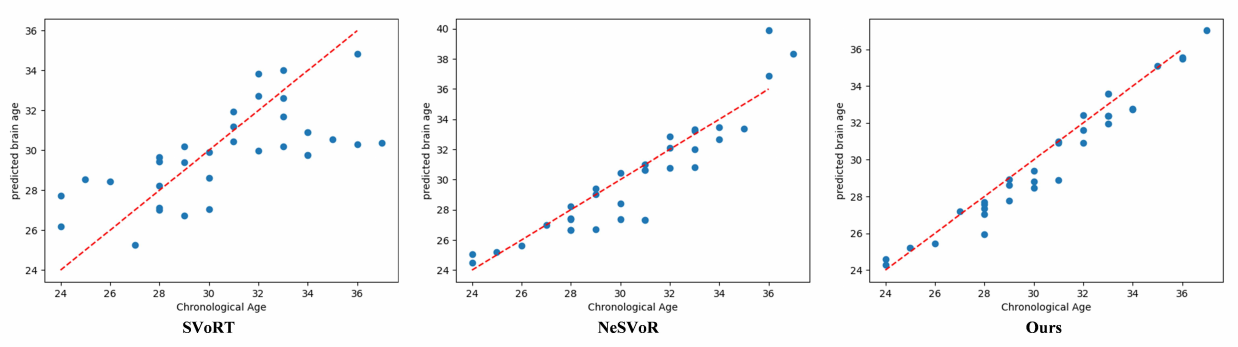}
\caption{The reconstructed fetal MRI results are validated for the downstream task of gestational age estimation on dataset A.} \label{fig8}
\end{figure}

In this part, we will quantify and analyze the gestational age estimation of the downstream task for the reconstruction results in Table 2 and Figs.~\ref{fig8}-\ref{fig10}, respectively. 

\subsubsection{Quantification of Fetal Brain Age Estimation:}

In Table 3, we show the reconstruction results of different algorithms for the downstream task of fetal brain age estimation quantified experimentally through the literature \cite{9446871}. As shown in Table 3, several common evaluation criteria are used, including the mean absolute error (MAE), the standard deviation (STD), Pearson's correlation coefficient (PCC), and Spearman’s rank correlation coefficient (SRCC). Notably, PCC and SRCC can be calculated between the estimated brain age and the chronological age, which is expressed as PCCage and SRCCage. On the other hand, the SRCC between the brain age gap and chronological age can also be calculated, denoted as SRCCgap. When MAE, STD, and SRCCgap are closer to 0, the algorithm shows better fetal brain age estimation results. However, when PCCage and SRCCage are close to 1, the reconstructed results show good results. According to the analysis, the reconstruction results of our proposed method achieve good results on the fetal brain age estimation task, both on the good original collection dataset A and on the poor quality original dataset B and large dataset C. According to the analysis, the reconstruction results of our proposed method achieve good results on the fetal brain age estimation task, both on the good original collection dataset A and on the poor quality original dataset B and large dataset C. 

Especially in dataset A, We randomly selected 12 samples to train the fetal brain age estimation network. The fetal brain age estimation effect of our proposed method is much better than the estimation effect of the other two methods. This shows that the fetal MRI reconstructed by our proposed method is of good quality under higher-quality acquisition data. Through the previous visualization results, the gray matter of our reconstructed fetal MRI volume is easier to distinguish, and the texture distribution above the white matter is more uniform. In dataset B, the MRI results reconstructed by NeSVoR are the least effective in estimating fetal brain age. Compared with the visualization effect of SVoRT, as shown in Fig.~\ref{fig3}, MRI reconstructed by NeSVoR has better results than SVoRT in gray matter and white matter, but the estimation effect of fetal brain age is worse than SVoRT. This is because the MRI volumes reconstructed by SVoRT in dataset B are brighter than those of NeSVoR, resulting in easier recognition by the fetal brain age recognition model. To verify this conclusion, we can also perform a comparative analysis by comparing the MRI reconstruction results of our proposed method in Fig.~\ref{fig3}. In the comparative analysis of SVoRT, NeSVoR, and the proposed RDGM, the reconstructed MRI volume from RDGM is brighter than the other two methods. In dataset C, SVoRT, NeSVoR, and the proposed RDGM are compared and analyzed, and their age estimation results are not very different. This is because dataset C is a large dataset containing 284 samples, and 242 samples were randomly selected for testing when performing fetal brain age estimation. Through the above analysis, our proposed MRI reconstruction method has better results in the validation of downstream fetal brain age estimation tasks.

\begin{table}[h]
\centering
\caption{The Results of Different Reconstruction Methods Are Tested for Fetal Brain Age Estimation on Three Datasetsand, and The Best Results Are Highlighted in Black.}\label{table2}
\resizebox{1\linewidth}{!}{
\begin{tabular}{c|c|ccccc}
\hline
\multirow{2}{*}{Datasets}  & \multirow{2}{*}{Methods} & \multicolumn{5}{c}{Evaluation Metrics}                                              \\ \cline{3-7} 
                           &                          & MAE            & STD            & PCCage         & SRCCage        & SRCCgap         \\ \hline
\multirow{3}{*}{Dataset A} & SVoRT                    & 2.003          & 2.47           & 0.683          & 0.766          & -0.659          \\
                           & NeSVoR                   & 1.311          & 1.399          & 0.907          & 0.706          & -0.634          \\
                           & Ours                     & \textbf{0.681} & \textbf{0.684} & \textbf{0.979} & \textbf{0.979} & \textbf{-0.157} \\ \hline
\multirow{3}{*}{Dataset B} & SVoRT                    & 3.545          & 4.028          & 0.591          & 0.613          & -0.892          \\
                           & NeSVoR                   & 5.686          & 5.302          & 0.306          & 0.405          & -0.921          \\
                           & Ours                     & \textbf{2.762} & \textbf{3.271} & \textbf{0.801} & \textbf{0.802} & \textbf{-0.873} \\ \hline
\multirow{3}{*}{Dataset C} & SVoRT                    & 2.871          & 3.113          & 0.799          & 0.774          & -0.895          \\
                           & NeSVoR                   & 3.243          & 3.293          & 0.705          & 0.794          & -0.793          \\
                           & Ours                     & \textbf{2.851} & \textbf{2.885} & \textbf{0.821} & \textbf{0.835} & \textbf{-0.784} \\ \hline
\end{tabular}}
\end{table}

\subsubsection{Scatter Diagrams of Estimated Brain Ages:}

\begin{figure}[t]
\centering
\includegraphics[width=0.5\textwidth]{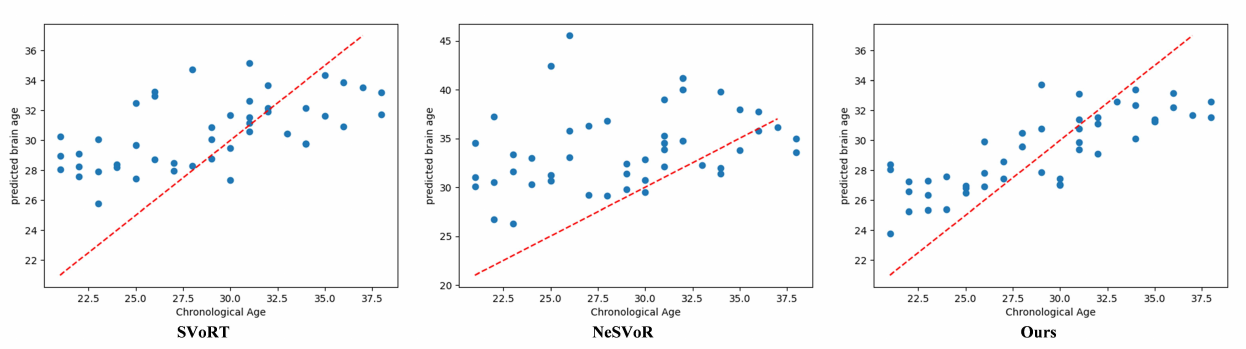}
\caption{The reconstructed fetal MRI results are validated for the downstream task of gestational age estimation on dataset B.} \label{fig9}
\end{figure}

\begin{figure}[t]
\centering
\includegraphics[width=0.5\textwidth]{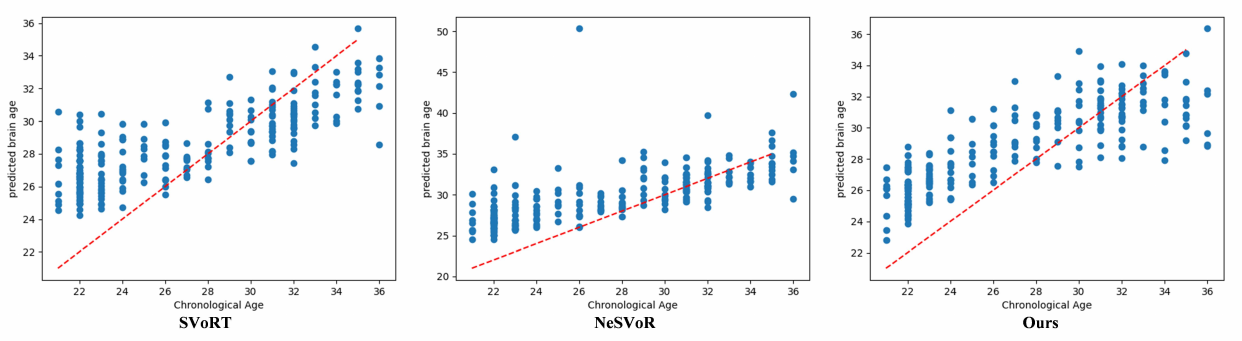}
\caption{The reconstructed fetal MRI results are validated for the downstream task of gestational age estimation on dataset C.} \label{fig10}
\end{figure}

As shown in Figs.~\ref{fig8}-\ref{fig10}, the distribution of gestational age prediction with linear regression is visualized for the three datasets. In Fig.~\ref{fig8}, the reconstructed MRI volumes by our proposed method have a more clustered scatter diagram in the downstream fetal brain age estimation task. However, SVoRT has the most scattered scatter diagrams. It can be proved that the MRI volume reconstruction effect of SVoRT fetus is not very good in distinguishing gray matter boundaries and white matter textures. This corresponds to dataset A in Fig.~\ref{fig3}. At the same time, we observe that the method NeSVoR is prone to outliers in gestational age estimation with large thresholds. This suggests that NeSVoR is deficient in processing the more complex gray and white matter facets. The MRI reconstructed by our proposed method is closer to the regression line, without outliers, and there are more points on the regression line than in the other two methods. This shows that in the high-quality collected data, our reconstruction effect is closer to the real-world data.

Compared with Fig.~\ref{fig8}, the scatter diagrams of Fig.~\ref{fig9} are not as good as those of Fig.~\ref{fig8}. According to the data quality analysis in Fig.~\ref{fig2}, the collection quality of dataset B is not as high as that of dataset A. Therefore, the overall reconstruction effect of fetal MRI of each method is not as good as that of dataset A. However, the scatter diagram of MRI in fetal brain age estimation reconstructed by our proposed method is still clustered near the regression line. Compared with the other two methods, the reconstructed results of our method are more robust in the regression line of fetal brain age estimation. The scatter diagram of SVoRT is more scattered around the gestational age of 22 weeks. It can be proved that SVoRT is not very effective in MRI volume reconstruction in the early fetal period. This corresponds to dataset A in Table 2. We observe that the method NeSVoR is prone to outliers in gestational age estimation around 26 weeks. This shows that our reconstruction effect can also be close to the real-world data in low-quality collected data.

In Fig.~\ref{fig10}, the estimated Scatter diagrams of fetal brain age for my dataset C are shown. Between 21-25 weeks of gestational age, the reconstructed MRI volumes of the three methods could not correctly predict brain age. And most of the correct points fall in the 26-36 weeks range. This indicates that the dataset has clear high-quality data acquisition in this range. Compared with SVoRT and our method, the regression lines of brain age estimation from reconstructed data are close. However, after careful observation, the gestational age estimation error of the SVoRT reconstruction data is larger than the brain age estimation error of the reconstruction data of our method in 21-25 weeks of the perinatal period. Most of the estimated gestational ages from the SVoRT reconstruction data are greater than 28 weeks within 21-25 weeks of the perinatal period. At the same time, at 34-36 weeks, the MRI reconstruction data of our proposed method are more clustered on the regression line. Notably, the gestational age estimates from NeSVoR reconstructed MRI data also have many outliers. By displaying and analyzing the scatter diagrams of each data set above, we can prove that the fetal MRI data reconstructed by our proposed method can obtain the optimal fetal brain age estimation effect.

\subsection{Convergence Analysis}

\begin{figure}[t]
\centering
\includegraphics[width=0.5\textwidth]{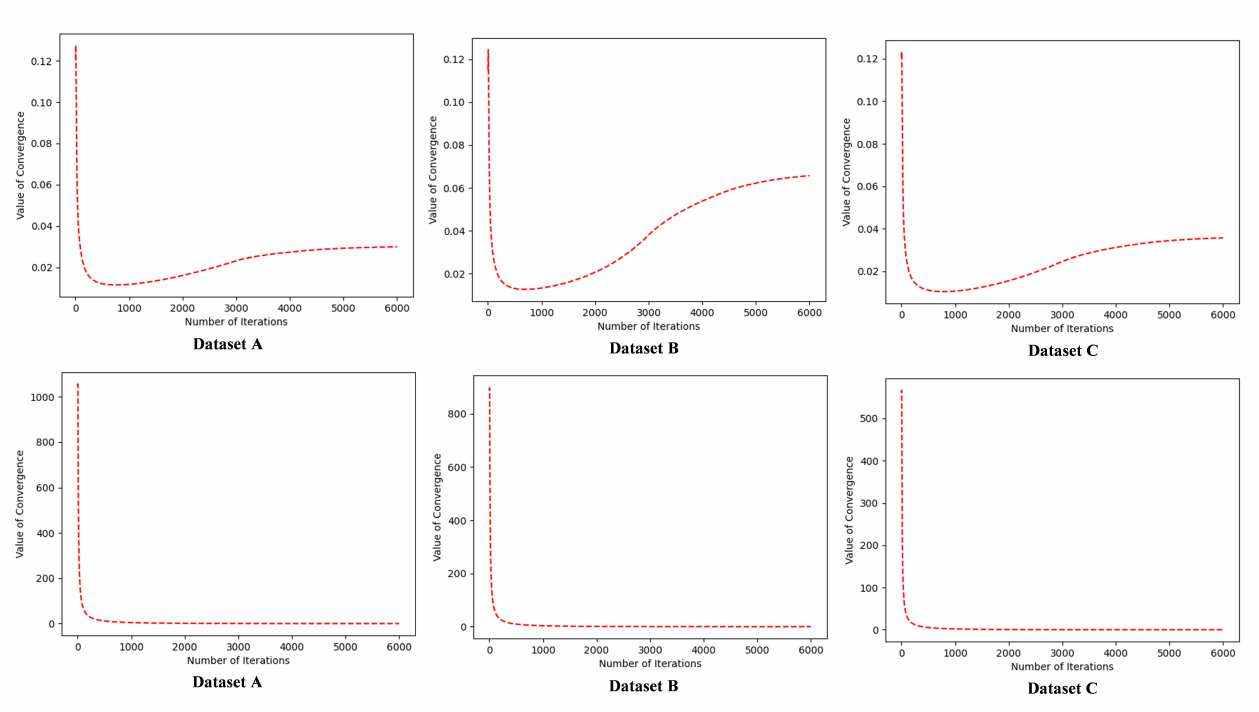}
\caption{The value of convergence results for different dataset samples. The first row shows the loss results of the method NeSVoR, and the second row shows the loss results of our proposed method.} \label{fig11}
\end{figure}

To further verify the performance of the proposed method, we implement algorithm convergence experiments on three data sets for sampling verification. As shown in Fig.~\ref{fig11}, we show the convergence curves of baseline NeSVoR and the proposed RDGM in three dataset instances. Among them, the first row in the figure is the convergence curve result of NeSVoR, and the second row is the convergence curve result of our proposed RDGM. Compared with the algorithm convergence curve of baseline NeSVoR, the convergence performance of our proposed method is better than NeSVoR. From the figure, it can be analytically concluded that our prominent method has a smoother convergence curve. Moreover, our proposed algorithm can quickly converge to 0 after hundreds of iterations. However, the samples of NeSVoR in the three datasets do not always show a decreasing trend and converge to 0. The convergence curve of NeSVoR presents a convex state. According to the analysis, our proposed method plays a great role in the processing of regional consistency. This not only makes the points in the sampling batch area more closely related, but also strengthens the overall structure of each sampling batch and the uniform distribution of pixels by optimizing the loss function.

\begin{figure}[t]
\centering
\includegraphics[width=0.5\textwidth]{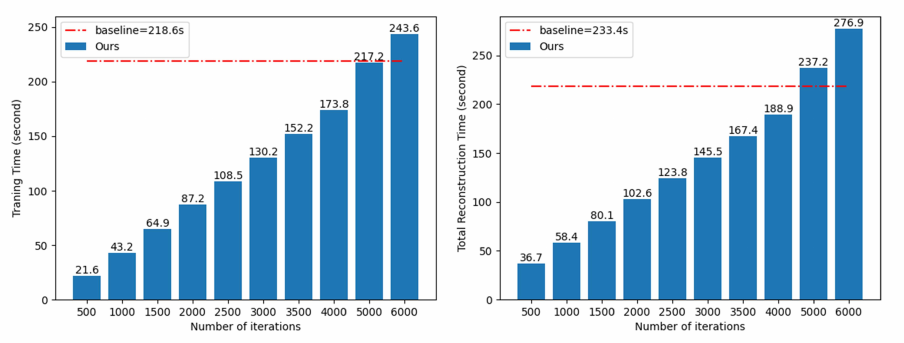}
\caption{The value of convergence results for different dataset samples. The first row shows the loss results of the method NeSVoR, and the second row shows the loss results of our proposed method.} \label{fig14}
\end{figure}

To further verify the reliability of our proposed algorithm, we analyze the running time of the algorithm in Fig.~\ref{fig14}. Our proposed method has more network results and super-resolution reconstruction mechanisms than baseline methods. This may cause more time consumption in the reconstruction of fetal MRI data. However, after the convergence analysis of our proposed method and baseline method in Fig.~\ref{fig11}, our proposed method can quickly achieve the convergence of the algorithm. Fig.~\ref{fig14} shows the running time for the three stacks input cases of our algorithm at different numbers of iterations. The dashed red line represents the time (6000 iterations) required by NeSVoR, the baseline method, to process one fetal MRI sample. The left side of the figure represents the time consumed by the self-supervised training of samples, and the right side of the figure represents all the time required for the reconstruction of samples. We analyze that in the sample training process when the number of iterations we use is 5000, the time consumed by our algorithm is less than NeSVoR. When both methods use the same number of training iterations, our method runs about 25 seconds more compared to the baseline. In terms of the total algorithm processing time comparison, our proposed method consumes 4 seconds more than the baseline at 5000 iterations. At 6000 iterations of training, our proposed direction consumes 44 seconds more compared to the baseline. Notably, in Fig.~\ref{fig11}, our proposed method has faster convergence results than NeSVoR. The number of training iterations is 5000 throughout the data processing. To do this, our approach adds a lot of mechanics and modules, but the runtime is about the same as the baseline.

\begin{figure}[t]
\centering
\includegraphics[width=0.5\textwidth]{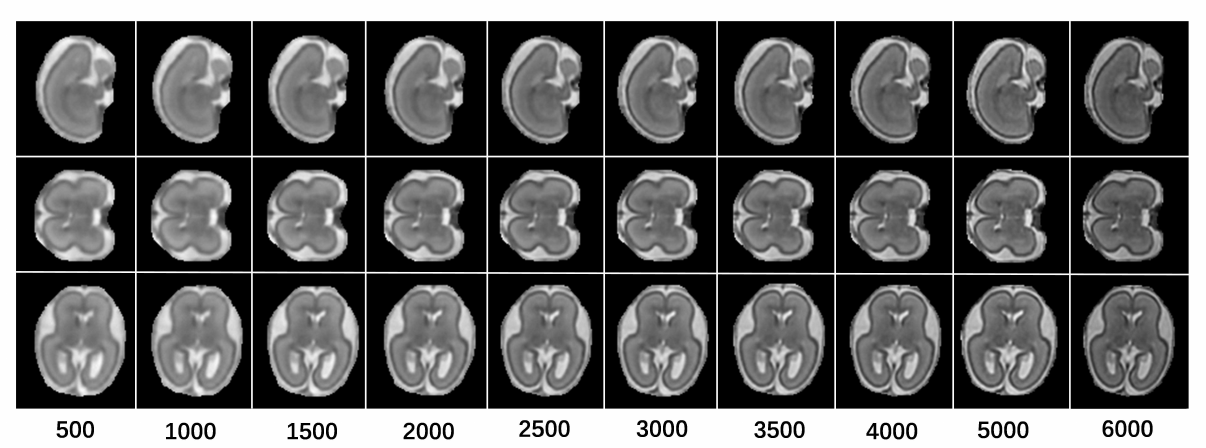}
\caption{The value of convergence results for different dataset samples. The first row shows the loss results of the method NeSVoR, and the second row shows the loss results of our proposed method.} \label{fig15}
\end{figure}

To further analyze the algorithm convergence and verify the selection of the number of training iterations, we visualize the reconstruction results for different numbers of training iterations. If as shown in Fig.~\ref{fig15}, we enumerate the reconstruction results for different numbers of training iterations in the range of 500 to 6000. As the number of training iterations increases, the results of fetal brain MRI reconstruction become better and better. When the number of training iterations reaches 4000, the reconstructed results of our algorithm are not too different from the reconstructed results of training iterations 5000 and 6000. This also verifies Fig.~\ref{fig11} that our proposed algorithm has a faster convergence curve. However, to ensure the reconstruction quality of all data, we self-supervised the training data using 5000 training iterations throughout the data processing.

\section{Conclusion}
In this work, we propose a new real-world multi-stack fetal brain MRI reconstruction model, called RDGM. This method incorporates slices rigid registration (SVORT), slice-to-volume regional consistency self-supervise generation (CINR), and volume-to-volume global associative super-resolution diffusion reconstruction (VDSG). For CINR, We propose a new NeRF network and achieve a regional coordinate association map of two different coordinate mapping spaces. To ensure the regional consistency of fetal brain MRI. We propose the robust region uniform distribution loss, which improves the slice reconstruction loss and image regulation loss for CINR self-supervised generation. By fusing slice reconstruction loss, image regulation loss, and bias field loss, the intensity uniformity distribution and robustness of fetal brain MRI rendering generation are strengthened. Combining CINR with the idea of a diffusion model, we propose a new VDSG mechanism, which is iteratively updated and generated to improve the quality of fetal brain MRI. The evaluation results show that RDGM not only generates more uniform regional and globally discriminated pixel intensity. RDGM also reconstructs motion artifact data obtained from various real scans well by self-supervised learning. Especially, our proposed method can achieve optimal reconstruction results in both high and low-quality acquisition data. In terms of algorithm convergence, it has a smoother and faster convergence curve. In the future, we will aim to investigate reconstruction algorithms for fetal MRI with fewer stacks and lower-quality acquisitions.

\bibliographystyle{elsarticle-num}
\bibliography{referrence}
\end{document}